# Modeling Morality

Walter Veit

University of Bristol

## Abstract

Unlike any other field, the science of morality has drawn attention from an extraordinarily diverse set of disciplines. An interdisciplinary research program has formed in which economists, biologists, neuroscientists, psychologists, and even philosophers have been eager to provide answers to puzzling questions raised by the existence of human morality. Models and simulations, for a variety of reasons, have played various important roles in this endeavor. Their use, however, has sometimes been deemed as useless, trivial and inadequate. The role of models in the science of morality has been vastly underappreciated. This omission shall be remedied here, offering a much more positive picture on the contributions modelers made to our understanding of morality.

**1** *Introduction*
**2** *Why model morality?*
**3** *Empirical Adequacy*
**4** *Implications for the Moral Status of Morality*

## 1 Introduction

Since Robert Axelrod's (1984) famous work *The Evolution of Cooperation*[1], economists, biologists, neuroscientists, psychologists, and even philosophers have been eager to provide answers to the puzzling question of why humans are not the selfish creatures natural selection seems to demand.[2] The list of major contributions is vast. Of particular importance is Brian Skyrms' pioneering use of evolutionary game theory (abbreviated as EGT) and the replicator dynamics in his books *The Evolution of the Social Contract* (1996) and *The Stag Hunt and the Evolution of Social Structure* (2004). Further important book-length contributions on the evolution of morality are offered by E.O. Wilson (1975), Ken Binmore (1994, 1998, 2005), Frans de Waal (1996, 2006), Elliott Sober and David Sloan Wilson (1998), Richard Joyce (2006), Jason McKenzie Alexander (2007), Martin A. Nowak and Roger Highfield (2011), Samuel Bowles and Herbert Gintis (2011), Christopher Boehm (2012), and most recently Patricia S. Churchland (2019).

---

[1] Based on an earlier co-authored paper with Hamilton 1981 of the same name.

[2] See Dawkins' 1976 book *The Selfish Gene* for an elegant illustration of the problem.

The efforts of these and many other authors have led to the formation of an interdisciplinary research program with the explicit aim to explain and understand human morality, taking the first steps towards a genuine science of morality. Let us call this research program the *Explaining Morality Program* (**EMP**). For a variety of reasons, models, such as those provided by Skyrms, have played a very important role in this endeavor, the most illustrative reason being the simple fact that behavior does not fossilize. Models and simulations *alone*, however, have been doubted by many to provide much of an explanation when it comes to human morality. The work of modelers in the **EMP** has been underappreciated for a number of reasons that can roughly be grouped together in virtue of the following three concerns: (i) the complexity of the phenomenon, (ii) the lack of empirical support, and perhaps the most threatening criticism being (iii) the supposedly non-reducible normative dimension morality embodies.[3] In this paper, I shall argue that this underappreciation is a mistake.[4]

Though interdisciplinarity has played a crucial role in the advancement of our moral understanding; it has led to an underappreciation of the role and contribution that highly abstract and idealized models have played. Many responses to the modeling work within the **EMP** are characterized by eager attempts to draw lines in the sand, i.e. determine prescriptive norms that would limit the justified use or misuse of such models.[5] These criticisms range from sophisticated ones, perhaps the most convincing one offered in Levy (2011) to rather naïve criticisms such as those offered in Arnold (2008). The latter goes so far as to label such models as useless, trivial and inadequate. In a harsh review of Arnold (2008), Zollman (2009) criticized Arnold's arguments against the use of models, deeming them unconvincing and exceedingly ambitious. Modelers may very well be tempted to attack Arnold as a straw-man and conclude that the criticism of models for the evolution can easily be debunked. However, such an approach would ignore the more sophisticated arguments that have been offered.

For the purposes of debunking the strongest arguments against such models, any mention of Arnold's (2008) criticism hardly deserves mention here. Arnold is by no means alone, however. His mistakes illustrate a shared pattern that can be found, though in a much weaker form, across the literature. It is a sort of a priori skepticism and perhaps dislike among philosophers and more experimentally oriented scientists, about the role of models in science. This skepticism is one I hope to at least partially dispose of here.[6] I shall demonstrate that models for the evolution of morality are neither too simplistic, nor do they lack empirical data to provide us with genuine explanatory insights. Recent advances in the philosophical literature on models, especially on model pluralism and the role of multiple models, should allow us to recognize not only such often exaggerated limitations but also the *strengths* of models in the **EMP**.[7] The latter of which have often been underappreciated, while the former have been overstated. This omission shall be remedied here.

---

[3] See Rosenberg & Linquist 2005; Northcott & Alexandrova 2015; Nagel 2012 respectively as examples for each.

[4] The EMP has faced similar criticism itself, relating not only to mathematical models but towards scientific explanations of morality at large. My goal here is only to provide a defence of the models used in this research program. I suggest, however, that if my attempt succeeds the entire EMP justifies its status as a genuine science of morality. Nevertheless, see FritzPatrick 2016 for a recent overview of EMP critics and defenders alike.

[5] See D'Arms 1996, 2000; D'Arms et al. 1998; Rosenberg & Linquist 2005; Nagel 2012; Northcott & Alexandrova 2015; Arnold 2008; Kitcher 1999; Levy 2011, 2018.

[6] Godfrey-Smith 2006, for instance, diagnoses a general distrust among philosophers in respect to "resemblance relations [of models] because they are seen as vague, context-sensitive, and slippery" (p. 733). Similarly Sugden 2009 has argued that models work by a form of induction "however problematic [that] may be for professional logicians" (p. 19).

In order to demonstrate a number of conceptual mistakes made in the literature, I shall largely draw on Jason McKenzie Alexander's (2007) book, *The Structural Evolution of Morality*, offering perhaps the most extensive modeling treatment on the evolution of morality. Building on previous work by his former supervisor, Brian Skyrms (1996; 2004), Alexander analyzes a large scope of exceedingly complex and arguably more realistic models, in order to illuminate the requirements and potential threats to the emergence and spread of moral behavior.[8] He concludes that morality can be explained by a combination of evolutionary game theory (abbreviated as **EGT**), together with a theory of bounded rationality and research in psychology. In doing so, he attempts to answer two distinct questions: (i) how could something as altruistic as human morality emerge and (ii) how did it persist against the threat of cheaters?

For the purposes of this paper, Alexander's (2007) book serves as a highly attractive case study for two reasons. Firstly, Alexander's contribution relies solely on highly abstract and idealized models of precisely the form often criticized as too simplistic to provide us with genuine insights for phenomena as complex as human morality. Secondly, while economists, psychologists, biologists, neuroscientists, and even political scientists have provided substantial contributions to the **EMP**, philosophers have offered distinct and extremely valuable insights by drawing conceptual distinctions.[9] As both a modeler and philosopher, Alexander treads very carefully only suggesting possible insights, his book may provide. At times, he even underestimates his own scientific contribution, arguing that it does not tell us much, if anything, without the supplementation of much more empirical data. One may regard such humility as a virtue, but at times, even an unbiased reader may get the impression that Alexander himself sees his contribution as superfluous. However, all of this humility seems to be thrown overboard in the very end of his book, where he discusses and suggests implications of the **EMP** for our understanding of morality itself, vindicating the 'objective status' of morality. Nevertheless, despite giving in too much to the criticisms of the program, Alexander avoids several pitfalls that might obscure our understanding of the epistemic contribution such models can provide. Here I shall shed a much more positive light on the role of models in the **EMP**, or as it sometimes referred to as: the study of *moral dynamics*.[10]

The structure of this paper corresponds roughly to the three concerns raised against the role of models in the **EMP** illustrated above. Firstly, in *Section 2,* I discuss Alexander's contribution and explore the most important question within the literature, i.e. why model morality? In *Section 3,* I respond to concerns regarding their empirical adequacy, before finally, in *Section 4,* I cast doubt on the possibility of vindicating the objective status of moral norms *via* the **EMP** and conclude the discussion.

---

[7] See Knuuttila 2011; Muldoon 2007; Wimsatt 2007; Weisberg 2007a, 2013; Ylikoski & Aydinonat 2014; Lisciandra 2017; Aydinonat 2018; Grüne-Yanoff & Marchionni 2018.

[8] To some extent one may treat his contribution as an extended robustness analysis of Skyrms' prior work. This would not do justice to Alexander's contribution, however.

[9] Richard Joyce's 2006 book *The Evolution of Morality* offers perhaps the most valuable contribution in this regard.

[10] See Hegselmann 2009.

## 2 Why model morality?

Evolutionary explanations of morality have been of scientific interest, since at least Darwin. Proto-Darwinian explanations, however, have been around for a long time. As Hegselmann (2009) points out, the **EMP** has a long scientific tradition going back as far as ancient Greece. Protagoras, in one of Plato's dialogues, provides perhaps the first scientific explanation of morality as a set of norms and enforcement agencies being an invention of humanity to escape a Hobbesian state of nature.[11] Couched in terms of a myth, we may treat this as a mere just-so story. Much later, David Hume came astonishingly close to providing a Darwinian explanation of morality himself.[12] Hegselmann and Will (2013) determine four key components to Hume's proto Darwinian account: a pre-societal human nature with confined generosity, the invention of artificial values to be reinforced and internalized through approval and disapproval of others, division of labour reaping the benefits of cooperation and trust and the "invention of central authorities that monitor, enforce, and eventually punish behaviour" (p. 186) already much more sophisticated but still similar to the myth of Prometheus and Epimetheus told by Protagoras. These accounts, a mere story and myth in the case of Protagoras and in the case of Hume an informal suggestion of a how-possibly explanation leave much to be desired, but they were, nevertheless, the best explanations available at the time. Luckily it didn't take two millennia for the next advancement. Richard Joyce (2006) in his book *The Evolution of Morality*, argues that "less than a century after Hume's death, Darwin delivered the means for pushing the inquiry into human morality further" (p. 228) filling out a gap Hume could only describe as *nature*. Charles Darwin, of course, himself suggested that the origin of morality can be explained with his theory:

> It may be well first to premise that I do not wish to maintain that any strictly social animal, if its intellectual faculties were to become as active and as highly developed as in man, would acquire exactly the same moral sense as ours. In the same manner as various animals have some sense of beauty, though they admire widely different objects, so they might have a sense of right and wrong, though led by it to follow widely different lines of conduct. If, for instance, to take an extreme case, men were reared under precisely the same condition as hive-bees, there can hardly be a doubt that our unmarried females would, like the worker-bees, think it a sacred duty to kill their brothers, and mothers would strive to kill their fertile daughters; and no one would think of interfering. Nevertheless, the bee, or any other social animal, would gain in our supposed case, as it appears to me, some feeling of right or wrong, or a conscience. (1879, p. 67)

This, of course, is still, 'just' a how-possibly explanation or as critics like to call them a just-so story. It should be clear that from Protagoras over Hume to Darwin, significant improvements in the explanation of morality have been made with more and more gaps being closed. Explanations come in degrees, and this research program is providing better and better explanations, and perhaps the best for the moment. Unfortunately, it took a while until informal evolutionary explanations resting on the good for the species were replaced with formal **EGT** models showing that the origin of moral behavior is not much of a mystery after all. Skyrms (1996) was the first to apply evolutionary game theory to unpack Hume's account in a formal manner, with others following in the creation of new models and simulations.[13] These sets of models strengthen our confidence that morality could have evolved in a way envisioned by Hume and Darwin, providing considerable explanatory power, even though empirical work has, hitherto, been largely left out of the picture. Even the work of moral philosophers in this Humean research program has been very empirical and guided by science trying to unpack the idea of morality being a mere artefact (see Mackie 1977; Joyce 2001, 2006) being a case in point for the division of labor between philosophers, modelers and empirical researchers. Modelers such as Skyrms simply continue an old philosophical school of thought with the modern tools of science, a move that ought to be encouraged.

---

[11] Plato. 1961. Protagoras. In: Hamilton E, Huntington C (eds) The collected dialogues of Plato. Princeton University Press, Princeton.

[12] Hume, D. 1998. An enquiry concerning the principles of morals (ed by Beauchamp TL). Oxford University Press, Oxford.

[13] See Alexander 2007; Hegselmann and Will 2010, 2013.

Before engaging in a more detailed analysis of our case study, i.e. the models Alexander (2007) provides, I shall take on his last chapter titled "Philosophical reflections" where he explores the philosophical implications of his models. Though the appearance of moral behavior in Alexander's models is rather robust and remains stable even in the face of defectors, more he argues needs to be said in order to draw inferences about human morality. Quoting Philip Kitcher, Alexander (2007, p. 267) highlights a general problem for evolutionary explanations of morality:

> [I]t's important to demonstrate that the forms of behaviour that accord with our sense of justice and morality can originate and be maintained under natural selection. Yet we should also be aware that the demonstration doesn't necessarily account for the superstructure of concepts and principles in terms of which we appraise those forms of behaviour. (Kitcher 1999)

In response, Alexander introduces a distinction between "thinly" and "thickly" conforming to morality. Though an individual's action may conform thinly with morality, e.g. fair sharing, the individual may fail "to hold sufficiently many of the beliefs, intentions, preferences, and desires to warrant the application of the term [']moral['] to his or her action" (2007, p. 268). In contrast, thickly conforming to morality satisfies sufficiently many of these conditions. If someone acts 'morally' out of purely selfish reasons, we may not want to call such behavior moral, e.g. someone giving to the poor in order to improve their reputation. Akin to Kant, it is the distinction between behavior in compliance with morality or acting out of the right, i.e. moral reasons.

When evolutionary game theory models are used to simulate the emergence and persistence of moral behavior we only observe the "frequencies and distribution of strategies and, perhaps, other relevant properties" (Alexander 2007, p. 270). What is lacking here is the role of psychology, perhaps even neuroscience, in the production of such moral behavior. Even if we allow for very complex strategies, such as those submitted to Axelrod's (1984) computer tournament, they still allow for a purely behavioral interpretation.

This problem lies at the core of attempts to model the evolution of morality. Critics argue that a complete explanation for the evolution of morality requires an understanding of the internal psychological mechanisms that produce such moral behavior. Alexander concedes to this criticism, suggesting to enrich these models with "non-strategic, psychological elements" (p. 273). He grants that **EGT** alone is not sufficient for an evolutionary explanation of morality, but that "together with experimental psychology and recent work in the theory of bounded rationality […] some of the structure and content of our moral theories" can be explained "by working in tandem" (p. 274). This position, of course, is a much weaker one than to claim that **EGT** *alone* could provide genuine insights into the origins of morality.

But even if, as I suggest, **EGT** might be sufficient to explain much of our moral behavior, Alexander aims at more. First of all, as Kitcher suggests, evolutionary game theory enables the important identification of behavior that maximizes long-run expected utility or fitness. A second step then is required to explain the motivational structures which are "actually *producing* this behavior in boundedly rational individuals" (2007, p. 275). Here Alexander identifies two mechanisms. First, the moral sentiments bringing about motivation to act, and secondly, moral theories instructing us "how to act once we have been motivated to do so" (275). Therefore, Alexander argues, it is precisely because we are boundedly rational that the "outcome produced by acting in accordance with moral theory are such that they tend to maximize our expected utility over the lifetime of the individual" (p. 275). Rationality requires us to rely on heuristics, and these luckily, according to Alexander, are often moral heuristics such as a fair split and cooperation.

In analyzing the influence of moral heuristics on our thinking, Alexander, based on a distinction by Sadrieh et al. (2001), discusses three separate roles that moral heuristics play in our thinking. Firstly, moral heuristics limit our set of options, e.g. not even considering to poison the dog of our neighbor, even though his barking may disturbs one's sleep. Secondly, moral heuristics guide our information search, i.e. what we need to consider before making a judgement. Thirdly, but closely related to the second point, moral heuristics "tell us when to terminate an information search" (p. 277). When we find out that someone killed a human infant for fun, it is sufficient for a moral judgement regardless of any additional information. Daniel Dennett (1996) has defended a similar position on moral judgements, calling them *conversation-stoppers* for otherwise costly debates. Relatedly, Alexander makes the rather contentious claim, that though we use moral reasoning, moral theories have their form precisely because they track "long-run expected utility" (p. 278). The key to the evolution of morality he argues "lies in the fact that we all face repeated interpersonal decision problems – of many types – in socially structured environments" (p. 278), hence the *structural* evolution of morality.

As "the science of morality is only in its infancy" (p. 281) there must remain some unanswered questions in our current explanation, however and I agree here with Alexander, this "is no reason why we should not make the attempt" (p. 282). Akin to primates, evolution equipped us with "emotions and other cognitive machinery" (p. 284), in order to solve interdependent decision problems, such as those arising in the prisoner's dilemma, the stag hunt, and the divide the cake game. Rosenberg makes a similar argument and extends it to love, as the "solution to a strategic interaction problem" (2011, p. 3). Analogously, the mere fact that love is an evolved response does not have to undermine our conviction that the feelings and intentions associated with it are not genuine or worthy of pursuit. The same may hold for morality. Emotions and our cognitive machinery is the raw material evolution had to use in order to solve more complex problems humans were increasingly facing, e.g. trust and the introduction of property rights.[14] With the evolution of language, this arms race in human evolution could only gain speed. We do not know yet which of our moral attitudes are hard-wired and which are culturally acquired, but that is obviously no reason not to ask the question. As we shall see, many of the EGT models used by Alexander, Skyrms, and others allow for both a cultural and biological interpretation. As cultural evolution operates at a much higher speed; however, many modelers such as Alexander (2007) give them a cultural interpretation.

Establishing the motivation for his models, Alexander henceforth, turns to the evidential support for evolutionary explanations of morality, in order to turn them into more than 'just-so stories', i.e. evolutionary explanations without empirical evidence. Evolutionary explanations are often faced with the criticism of providing nothing more than 'just-so stories', i.e. historical accounts without any empirical evidence in their favour. For Charles Darwin, it was very important to collect plentiful evidence for his theory of natural selection and biologists to this day continue to accumulate corroborating evidence. However, when biologists try to explain the occurrence of a certain behavior or a phenotype in general, they often start by hypothesizing how the trait could be adaptive. This research program is often criticized as a sort of *Panglossian adaptationism*, i.e. assuming the adaptiveness of a trait without further evidence.[15] Though Alexander only considers two experiments, (see Yaari and Bar-Hillel 1984; Ken Binmore et al. 1993), they are highly suggestive that our conception of fairness is somewhat flexible and strongly correlates with the outcomes our own group receives. Though a philosophical review of the vast literature on moral experiments should be undertaken, it is beyond the scope and purpose of this paper, which is merely concerned with attempts to model morality.[16] Nevertheless, the just-so story critique has evolved into a term of abuse used against all kinds of model-based explanations. In this paper, I shall attempt to argue against this commonplace treatment and highlight the wealth and diversity models can provide in the **EMP**.

---

[14] Here the often drawn distinction between biological and psychological altruism plays an important role.
[15] See Gould and Lewontin 1979 for their famous critique of adaptationism.

Though brought up as a game theorist, Alexander recognizes the weakness in the assumptions of standard rational choice theory. In order to avoid charges of unrealisticness, Alexander's models for the evolution of moral behavior make no strong rationality assumptions; rather he uses models of bounded rationality combined with evolutionary game theory to account for the evolution of morality. Sugden anticipated as much in a paper on the evolutionary turn in game theory stating that the "theory of human behaviour that underlies the evolutionary approach is fundamentally different from that which is used in classical game theory" (2001, p. 127), with far less contestable rationality assumptions, though similar in their mathematical formulation. In short: Alexander treats bounded rationality theory as descriptively superior to standard rational choice theory.

However, with the threat of only providing so-called *just-so stories*, evolutionary explanations, in general, are often dismissed by pointing to the multiplicity of evolutionary accounts we could give for the appearance of a phenomenon. These objections, however, miss the mark when they supposed to show that evolution plays no part in explaining morality. Alexander's former supervisor, Brian Skyrms, himself working on the evolution of social norms makes this criticism of just-so story charges explicit: "Why have norms of fairness not been eliminated by the process of evolution? […] How then could norms of fairness, of the kind observed in the ultimatum game, have evolved?" (1996, p. 28). In this section, I argue that such criticism is highlighting something important that Robert Sugden (2000; 2009) tries to capture in his work on model-based explanations. Though very similar arguments have been made by Giere (1988, 1999), Godfrey-Smith (2006), Weisberg (2007b, 2013), and Levy (2011), Sugden's work serves as an elegant illustration of Alexander's aims for at least two reasons: (i) Sugden's account is partially motivated by evolutionary game theory models used in both economics and biology, and (ii) his 'credible world' terminology maps neatly onto the justifications, goals, and inferences Alexander is drawing himself.

Models, Sugden (2000, 2009) argues are parallel worlds, artificially created, which can be used to draw inductive inferences to the real world. At least he argues, such is the practice in economics and biology. In both of these fields, phenomena are complex and can be multiply realized by different mechanisms. This is why, Sugden argues, we need induction to bridge the gap between the model world and the real world, even though he grants that this may seem unappealing to some philosophers. A model here, in virtue of its idealizations, is a sort of fictional entity that enables us to draw inductive inferences about the real world via similarity relations to the 'model world'. Hence, Sugden argues, modelers aim to create 'credible worlds' that we *could imagine being real*. It is not truth *per se* that is aimed for, but rather a sort of credibility that is deemed able to tell us something about the real world we live in. To do so modelers, are required to provide us with relevant similarities between what is happening in the model and what could be going on in the real world, perhaps requiring a sort of elaborative story or narrative linking the two. In the following, I argue that Alexander's contribution to the **EMP** consists in the construction of such 'credible worlds' from which we can draw inductive or abductive inferences to the real world.[17]

Analyzing the phenomena cooperation, trust, fairness and retribution, Alexander (2007) conducts his project by exploring different and increasingly complex models in which he wants to explore the evolution of morality.[18] He goes on to employ five models, i.e. replicator dynamics, lattice models, small-world networks, bounded-degree networks and dynamic networks each introducing more and more elaborate forms of population structure back into the picture and increasing the realism of his models. He analyses four different dimensions of morality, i.e. cooperation, trust, fairness and retribution amounting to a set of twenty models, each having their robustness tested in several iterations. Each of these models alone seems to tell us very little about the real world. Taken together, however, this extensive set of robust models supports Alexander's assertion that population structure plays a very important role in the evolution of morality.

---

[16] See Kagel, J. H. & Roth, A. E. 1998 for an overview of such studies in experimental economics.

[17] I treat abduction, i.e. inference to the best explanation, here similar to Sugden 2009 as a form of induction. Others do not share this view, instead arguing that eliminative induction is a form of IBE e.g. Aydinonat 2007, 2008. However, I have no bone to pick in this debate. What conception one holds does not impact the validity of the arguments presented here.

First, he starts with a simple model used in evolutionary biology and increasingly the social sciences, i.e. the replicator dynamics. As already alluded to, **EGT** allows for both biological and cultural interpretations explaining the interdisciplinary interest in **EGT**. While the biological form of these models treats replication as (biological) inheritance, replication has to be interpreted as some form learning or imitation in a cultural setting. Replicator dynamics (RD) are an attempt to model the relative changes of strategies in a population. Again, these can be either instantiated biologically or culturally. Strategies with higher fitness than the population average prosper and increase their share in the population, while those with lower fitness are driven to extinction. RD in the biological setting are thus an attempt to model the dynamics of reproduction and natural selection. The following is the continuous replicator dynamics equation:

$$\frac{dx_i}{dt} = [u(i,x) - u(x,x)] * x_i \qquad \text{(Weibull, 1995, p. 72)} \qquad (1)$$

In each round individual strategies, i increase their share within a population linear to their success u(i,x) compared to the average fitness u(x,x) in the population. Just as the evolutionary stable strategy (ESS) familiar from earlier evolutionary game theory models, RD assume infinite population size or at least infinite divisibility and random interaction. These idealizations allow us to analyze the frequency-dependent success of different strategies, whether they are biologically or culturally transmitted. Though he intends his project to model the cultural evolution of morality, he grants that replicator dynamics leave it open whether the strategies are genetically or culturally transmitted. Let us consider Alexander's first example and the most-analysed game in game theory: the *Prisoner's Dilemma*.

Table 1 *The payoff matrix for the Prisoner's Dilemma*

|                      | Cooperate | Defect |
|----------------------|-----------|--------|
| Cooperate (Lie Low)  | *R*       | S      |
| Defect (Anticipate)  | T         | P      |

In the *Prisoner's Dilemma,* there is only one NE, i.e. Defect, Defect. This famous game can be traced back to Hobbes (1651), who argued that a powerful leader is required to escape the state of nature, i.e. collective defection. In fact, his name is mentioned over twenty times in

---

[18] See Gelfert 2016 for a recent discussion of the various exploratory functions of models.

Alexander's book, pointing to the long tradition of the **EMP**. In Table 1, T is "temptation", i.e. the value tempting defection, R is the "reward" of joint cooperation, P is "punishment" as both receive a lower payoffs then they would have gotten if both had cooperated, and S is the "sucker's" payoff where a co-operator is exploited (2007, p. 55). The payoffs are ordered as follows T > R > P > S with the additional condition that T + S/2 is smaller than R. The ESS here coincides with the strict Nash Equilibrium (**NE**)[19] predicting mutual defection. Using replicator dynamics to model the evolutionary trajectory shows that co-operators are quickly driven to extinction, with defectors taking over the population.

As his book is called *The Structural Evolution of Morality*, Alexander is aware that human societies are more complex and that we need to account for the social structure of society in order make these models more *credible*. In fact, when population structure is introduced, and interactions are no longer entirely random, making it possible for co-operators to group together, cooperation can persist and evolve. Therefore he moves on to explore agent-based models, i.e. lattice models, small-world networks, bounded-degree networks and dynamic networks, where agents can choose with whom to interact. Increasing the complexity in his models serves then two purposes: on the one hand the goal is (i) to ensure robustness, i.e. the stability of the outcomes in the model under changes in the model and on the other hand (ii) to increase the credibility of the model, i.e. the likeliness of it telling us the true story about the evolution of morality. As Sugden says: "what we need in addition is some confidence that the production model is likely to do the job for which it has been designed – that it is likely to explain real-world phenomena" (2000, p. 11), and this is Alexander's stronger aim: the prevision of a *how actual* explanation.[20] Let me therefore, now tackle these two purposes in succession.

Looking at robustness first, Alexander claims that the results in his models are sufficiently robust to suggest that moral behavior can emerge and remain stable in a population of boundedly rational agents. I agree with Kuorikiski and Lehtinen (2009) that robustness analysis is somewhat implicit in Sugden's account of inductive inference[21], this, however, should be interpreted as a continuous inference from increasingly similar model worlds to their real-world counterpart. Robustness analysis and inductive inference are closely related and overlap in important respects, but Sugden is justified in making a distinction on the grounds of their different epistemic properties. Robustness analysis increases internal validity for the model world, while this internal validity is a prerequisite for establishing external validity in the real world. When slightly changed alterations of the model are seen as the target themselves, this distinction breaks down. I take Levy's (2011) subtle criticism on the modeling literature on the evolution of morality, to target the tendency of not sufficiently distinguishing between these two distinct ways in which validity can be increased. Modelers such as Skyrms, Levy argues, take their models to establish external validity, when really only internal validity has been vindicated. I will say more about Levy's criticism in the next section on the empirical adequacy of these models.

When models are used to learn about human morality, Sugden (2009), Cartwright (2009) and others are correct in arguing that the purpose of models is to learn something about a real-world target system. Francesco Guala, argues that it is "necessary to investigate empirically which factors among those that may be causally relevant for the result are likely to be instantiated in the real world but are absent from the experiment (or vice versa)" (2005, p. 157). This procedure of establishing external validity not only applies to inductive inference from the artificial experimental world to the real world but also to inductive inference from the artificial model world to the real world. In both cases, we want to draw inferences from highly idealized and abstract mechanisms to a causal mechanism operating in the real world. As several authors have recently pointed out, there are more important similarities than relevant differences between models and experiments, which makes it difficult to justify drawing any hard boundaries between the two.[22]

---

[19] The Nash Equilibrium introduced by John Nash, is the central, in fact, most important solution concept in Game Theory. The concept picks out a combination of strategies, i.e. one for each 'player' in the game, in which none of the players has an incentive to unilaterally deviate from his chosen strategy, while the strategies others have chosen remain fixed. In the Prisoner's Dilemma this classically leads to only one unique solution, i.e. mutual defection. Morality quickly suggests itself as an evolved social solution to such inefficient equilibria.

[20] I use how actual explanations in a modal sense, i.e. as a subset of how possibly explanations.

[21] One may even treat robustness analysis as a necessary component of model-based science itself. Sometimes it is used in a very narrow sense, at other times quite broadly. See Lisciandra 2017 for a recent overview, but also Woodward 2006.

In order to gain confidence that Alexander's 'story' provides us with the actual explanation of human morality, requires more, especially evidence from psychology and neuroscience, in order to learn about the causal mechanism behind moral behavior. Even though our models are robust, this robustness in itself only tells us something about the evolution of moral behavior in the model, i.e. unless relevant similarities obtain between the real and the model world. Sugden argues that "a transition has to be made from a particular hypothesis, which has been shown to be true in the model world, to a general hypothesis, which we can expect to be true in the real world too" (2000, p. 19), i.e. inductive inference. Sugden explicates three such inductive schemata, explanation, prediction and abduction. For the purpose of this paper, only his explanation schema is relevant:

E1 – in the model world, R is caused by F.
E2 – F operates in the real world.
E3 – R occurs in the real world.
Therefore, there is a reason to believe:
E4 – in the real world, R is caused by F. (2000, p. 12)

The phenomena R in question is the emergence and stability of moral behavior. Though Alexander explicitly wants to explain more, i.e. the emergence and stability of *morality*, we shall first consider whether these models can explain 'moral' behavior. What we need to establish in order to make justified inductive inferences, i.e. extrapolation, from the model world to the real world are relevant similarities. While the relevant set of causal factors in the model, i.e. cultural evolution, do operate in the real world this may be an unavoidable feature of generalized theories. When Sugden speaks of a model's *credibility* he is not talking about their literal truth, but *truthlikeness*, a description of "how the world could be" (2000, p. 24), a credible counterfactual world. For a model world to achieve this kind of credibility, it needs to cohere with the causal processes we know to be operating in the real world. The agents postulated in our model need to be in a relevant sense like real agents in our world. By using evolutionary game theory and bounded rationality models, Alexander intends to trump the standard rational choice theory models in virtue of credibility. Drawing inferences from his models, therefore, he argues is at least inductively more justified than standard game theory explanations for the evolution of morality. This concession, however, is unlikely to convince many critics of rational choice theory.

However, if standard rational choice models are justified, his models should be justified by extension in virtue of their enhanced credibility. If the standard models fail to achieve credibility or rather external validity, we need some further argument, to see how Alexander's models are explanatory. When Alexander moves from the simple replicator dynamics to lattice models, he is continuing his quest for a more credible model world. Here we drop the unrealistic assumption of random interaction in an infinitely sized population for a one-dimensional lattice in which everyone has two neighbors to interact with. Secondly, Alexander analyzes how different learning rules change the strategy dynamics in his models, all of which are rather simple but perhaps better capture the actual strategy changes in human agents. As the assumption of only interacting with two neighbors is highly unrealistic in itself, Alexander moves to small-world networks where some agents have an additional interaction possibility by being connected over a bridge. Further increasing credibility, Alexander moves to bounded-degree networks where every agent has a certain number of connections $i$ between k(min) and k(max). Here connections need not be neighbors and are randomly assigned, creating networks that look fairly similar to interaction networks in the real world. However, humans obviously do not choose with whom to interact entirely at random. When we encounter someone who cheats in a cooperative endeavor we will try to avoid them and interact with someone else next time. Alexander draws on Skyrms' & Pemantle's (2000) model of social network formation in order to model changing interaction frequencies. Without going into the specifics and intricacies of each of these models, they illustrate an important point: Alexander's book follows the modeling strategy of first ensuring robustness and internal validity, before moving on to more credible model worlds that gain complexity and inferential power. The latter approach must, of course, be closely related to empirical research into morality, most importantly, perhaps moral psychology.

---

[22] See Mäki 2005 and Parke 2014.

Robert Sugden's account of modeling is justified in virtue of being a naturalistic, pragmatic account of the actual scientific modeling practice. Models are successful in explaining but do so by induction. Therefore, we should accept induction as a valid principle in the modelers toolkit, "however problematic [that] may be for professional logicians" (2009, p. 19). In a research paper on the evolutionary turn in game theory, Sugden writes:

> Evolutionary game theory is still in its infancy. A genuinely evolutionary approach to economic explanation has an enormous amount to offer; biology really is a much better role model for economics than is physics. I just hope that economists will come to see the need to emulate the empirical research methods of biology and not just its mathematical techniques. (2001, p. 128)

Alexander doesn't fall into this trap, for he only sees his mathematical models as a subset of the necessary steps towards a genuine explanation for the evolution of morality. This is where empirical evidence needs to be accumulated, and studies conducted, analyzing development psychology with respect to social norms. Alexander's work, however, guides the way for such empirical research and theory testing to commence, in a field that is still nebulous and wide. How to move from robust EGT models to the real world will be explored in the next section.

## 3 Empirical Adequacy

The most sophisticated criticism of attempts to model the evolution of morality has recently been offered by Arnon Levy (2011). Rather than denying the importance of models, Levy argues that there are two distinct modes of inquiry in which indirect modeling can be used to study otherwise complex phenomena. Levy argues that the work of Skyrms, Alexander, and others is characterized by 'internal' progress, achieved within the model, rather than 'target-oriented' progress where we learn more directly about the target system itself. While Levy does not go as far as to argue that this strategy is pure conceptual exploration[23], he suggests that it is "more conceptual in spirit" aimed at understanding the initial model itself (2011, p. 186). Target-oriented modeling, Levy argues, progresses by "incrementally adding causal information" primarily guided by considerations of empirical adequacy (p. 186). In contrast, models for the evolution of morality explore the "subtleties of a constructed set-up" with empirical adequacy only playing a minor role (p. 186).

---
[23] See Hausman 1992.

Similarly, Sugden suggests that a "model cannot prove useful unless someone uses it, and whoever that person is, he or she will have to bridge the gap between model world and real world" (2009, p. 26). Though Alexander downplays the role of his models by saying that they alone cannot account for much, he suggests that jointly with theories of bounded rationality and research in psychology and economics we can get closer to the actual explanation of how morality evolved. This is the main motivation behind Sugden's (2000, 2009) credible world account of modeling. If models were only about conceptual exploration and providing theorems, any mention of the real world and the relationship of the model to it would be *nothing but* telling a story to sell one's model. Levy (2011) suggests that this is what might be happening in the models provided by Skyrms and Alexander. However, as I shall argue, their modeling strategy of the evolution of moral norms explicitly acknowledges relevant real world factors and successively tries to increase the credibility of his models. In the following, I shall argue that **EGT** models can inform empirical research and vice-versa.

Rosenberg and Linquist (2005) wrote a paper on evolutionary game theory models for cooperation and how to test them empirically, which will prove highly useful for this section. They argue that we can use archaeology, anthropology, primatology and even gene-sequencing to support **EGT** models. Supporting Alexander, they argue that human cooperation is too sophisticated, conditionalized and domain-general for it to be a genetically hardwired trait. What Alexander does not consider is potential gene-culture co-evolution. But I take him to make a deliberately weaker claim, that even if it turns out that human morality is entirely cultural, his models will be useful. If we find empirical evidence for 'hard-wired' behavior than all the better for an evolutionary explanation of morality. However, in order to have a credible **EGT** model Rosenberg and Linquist argue already requires a lot of substantial assumptions, for example, emotion, reliable memories, a theory of (other) mind(s), language and imitation learning. Experiments in economics and psychology, often done in the form of games, can inform us about how humans act, and how a change in conditions changes human behavior. Such empirical work can then help the modeler to not only increase the credibility of his models but also to eliminate those models would tell a completely misguided story of the evolution of (human) morality.[24]

Rosenberg and Linquist provide the popular example of the big-game hunt hypothesis as an explanation for why humans started cooperating. They point out that the empirical data suggests that big-game hunter was an inferior strategy in comparison to gathering, not even granting the payoffs specified in a stag-hunt game. Though the big-game hunt hypothesis tells a nice story of why humans started to cooperate, we should treat it as even less than a 'just so story' because the evidence suggests that it is most likely false. As an alternative, they propose cooperative child-caring which interestingly also fits the mathematical description of a stag-hunt game. In light of empirical evidence, they argue that modelers should try to alter their stag-hunt models for trust by thinking about the potential payoffs of cooperative child-care rather than the payoffs of cooperative hunting.

Such is the nature of this enterprise: both modeling and empirical research can inform each other in a variety of ways. Though many models will be discarded, this leaves us with a much narrower set of how possible explanations and gets us closer to the actual one. They close their paper by stating that is "not for philosophers to speculate how this research once commenced will eventuate" (p. 156), but it is nevertheless necessary to bring in line the theoretical work done in evolutionary game theory with the empirical data from various field.[25] Otherwise, **EGT** models are nothing but conceptual exploration, and as Sugden points out, modelers should and can aim for more. Nevertheless, Hegselmann (2009) suggests caution against the hope that the "huge gap between macroscopic models of moral dynamics and the known variety of microscopic processes that seem to generate certain assumed overall effects" (p. 5689) can be bridged in the near future, if at all. Criticism directed against Alexander, stating that he failed to provide a complete explanation is not nearly as effective when a complete explanation is not reachable anyway. While we might consider Alexander's work as only one of the first steps in getting closer to a complete explanation of human morality, it remains an important step nonetheless. Though the empirical data is weak, or rather because of it, it is so important to combine the research results from different fields. Models, far from being a mere add-on to this research program, appear to be a necessary and integral part of the **EMP**, transforming it more and more into a science in its own right. Let me now conclude this discussion with the controversial debate whether such models have any impact on the *moral* status of morality.

---

[24] As Zollman 2009 points out in his review of Arnold 2008, models have directly inspired experimental work on the evolution of morality, e.g. Wedekind and Milinki 2000; Seinen and Schram 2006.

## 4 Implications for the *Moral* Status of Morality

Alexander concludes that "evolutionary game theory, coupled with the theory of bounded rationality and recent experimental work bridging the gap between psychology and economics, provides what appears to be a radical restructuring of the foundations of moral theory" showing that the content of "moral theories are real and binding" (2007, p. 291), though their content is highly dependent on us and the structure of society. Alexander (2007), rather than providing an evolutionary debunking argument for morality, claims to provide an 'objective but relative' basis for morality in so far as he shows that the principles of morality are in the best long-term interest of everyone, a claim that may seem just as radical. This distinction is important as it is often treated anonymously; however, as I shall argue, Alexander goes one step too far when he treats the instrumental justification of morality as an epistemic justification.[26]

Due to the importance of population structure illustrated in his book, Alexander argues morality is necessarily relative to the structure of society. Morality, he argues, is objective but relative. This is an ambitious suggestion, standing in stark contrast to the careful conclusions Alexander has drawn in the rest of his book, and hence deserves closer inspection. Unlike Joyce (2001, 2006), Street (2006) and Rosenberg (2003, 2011), who provide evolutionary debunking arguments for the objectivity of morality, Alexander argues that his models, rather than undermining, are able to vindicate morality. And in doing so, he draws explicitly on Hume:

> [M]uch work has to be undertaken in order to unpack Hume's "certain proposition" that "[T]is only from the selfishness and confin'd generosity of men, along with the scanty provision nature has made for his wants, that justice derives its origin."[27] And, as for the origin of justice, so for the rest of morality. (2007, p. 291).

For Humeans in the tradition of Bernard Williams (1981) and arguably Hume himself, notions of objectivity were always somewhat odd. In fact, all of the three evolutionary debunkers mentioned above, see themselves as Humeans. They argue that in light of an evolutionary explanation for the adaptiveness of moral attitudes and behavior, there is neither a need nor should we endorse any 'magical' property that makes morality somehow objective. Nevertheless, Richard Joyce (2001) in his book *The myth of morality* explores the possibility of vindicating the objectivity of morality by linking it to rationality, as perhaps the strongest candidate view to avoid the conclusions of the error theorist (see Mackie 1977). Alexander's argument for the vindication of morality rests on the same motivation: if it can be shown that it is in everyone's interest to act according to morality, morality can be saved.

---

[25] Similar arguments have recently been raised against the use of game theoretic tools to explain the evolution of multicellularity. See Veit 2019a.
[26] I thank Richard Joyce for suggesting this formulation to me.
[27] *A Treatise of Human Nature*, Book III, part II, section II.

For this approach to be successful, Joyce (2001) argues we would need to arrive at some sort of categorical imperative that derives from rationality alone, i.e. precisely the route Kant took to save the status of morality from Hume's philosophy. For Humeans, who see reasons as relative to desires, aims and preferences (perhaps also beliefs), this approach must be futile. The moral heuristics will not only be relative to the structure of the society we live in, but also relative to the aims and desires we have, and hence subjective. They would be nothing more than mere heuristics that apply to the majority of the population in the majority of circumstances. Alexander does not see this as a problem; in fact, he sees it as sufficient for grounding morality as something objective, but nevertheless relative.[28]

However, as I see it the previously provided arguments are sufficient for casting doubt on the project of vindicating the objectivity of morality by pointing to a highly relativistic notion of rationality crucially depending on the social structure of society.[29] For even if we grant that this is a sort of objectivity, it is not what humans refer to when talking about the objectivity of morality, nor is it what metaethicists are usually interested in. Error theorists like Mackie, Joyce and Rosenberg readily accept the debunking. Alexander, however, prefers a more subtle version of what we could mean by moral objectivity. His work captures something important: the advice we give our children, the moral norms we teach are likely to be in their long-term best interest. They are useful heuristics that evolved to reap the benefits of cooperation in strategic interaction problems, and as Alexander points out, they are highly contingent on the social structure of society. Levy (2018) suggests that the models explored in the **EMP** could provide us with insights into the desirability of certain institutions and societal norms, merely in virtue of their stability. For meta-ethics, however, the impact of the **EMP** may be severe. Akin to an electromagnetic pulse (**EMP**) the Explaining Morality Program could paralyze much of the traditional work of philosophers working on morality.[30] Hence, it comes with no surprise that many naturalists and philosophers of science seem to hold a deflated sense of moral objectivity or become error theorists, such as Mackie.

Much more empirical work needs to be done, but the long path to explaining morality is at least partly illuminated by the work of modelers such as Skyrms, Alexander, and others. Clearly, this can only be the beginning of an explanation, but the first steps have been taken. Replicator dynamics have limits and often need to be supplanted with other models, e.g. non-**EGT** models for inheritance and cognitive mechanisms, to provide satisfying explanations of real-world phenomena. A diverse set of multiple models among which as Hegselmann (2009) argues bridges can be built may be the best thing we can hope for, but these as I have argued are importantly not without considerable explanatory power. It is just the faulty ideal of a complete explanation that blocks such incremental steps towards a better understanding of complex phenomena such as human morality.

---

[28] Richard Joyce 2001, 2006 explores these issues in more detail than I can do justice here.

[29] Sterelny and Fraser 2016 offer a defence of such a weaker form of moral realism. I will note that I do not find such approaches plausible, as they commonly rest on a re-definition of what is traditionally understood as moral realism.

[30] This, however, is a matter for another paper. Nagel 2012 recognizes this threat but turns the modus ponens into a modus tollens even going so far as to argue that since moral realism is true the Darwinian story of how morality evolved must be false. This gets things backwards. See Garner 2007 for radical conclusions regarding the elimination of morality, or for nihilism more generally, see Sommers & Rosenberg 2003 and Veit 2019b. A collected volume on the question whether morality should be abolished has recently been published by Garner and Joyce 2019.

**Acknowledgements**

First of all, I would like to thank Richard Joyce, Rainer Hegselmann, Shaun Stanley, Vladimir Vilimaitis, Gareth Pearce and Geoff Keeling for on- or offline conversations on the topic. Furthermore, I would like to thank Cailin O'Connor, Caterina Marchionni, and Aydin Mohseni for their comments on a much earlier draft that was concerned with evolutionary game theory models more generally and Topaz Halperin, Shaun Stanley and two anonymous referees for comments on the final manuscript of this paper.

Also, I would like to thank audiences at the 11th MuST Conference in Philosophy of Science at the University of Turin, 2018's Model-Based Reasoning Conference at the University of Seville, the 3rd think! Conference at the University of Bayreuth, the 4th FINO Graduate Conference in Vercelli, the Third International Conference of the German Society for Philosophy of Science at the University of Cologne and the 26th Conference of the European Society for Philosophy and Psychology at the University of Rijeka. Sincere apologies to anyone I forgot to mention.